\documentclass[twocolumn,aps,amsmath,amssymb,letterpaper,superscriptaddress]{revtex4-1}

\usepackage{graphicx}
\usepackage{amsmath,amssymb,amsfonts}
\usepackage{braket}

\begin{document}

\title{Correlated Spin-Flip Tunneling in a Fermi Lattice Gas}
\author{Wenchao Xu}
\affiliation{Department of Physics,  University of Illinois at Urbana-Champaign, Urbana, Illinois 61801, USA}
\author{William Morong}
\affiliation{Department of Physics,  University of Illinois at Urbana-Champaign, Urbana, Illinois 61801, USA}
\author{Hoi-Yin Hui}
\affiliation{Department of Physics, Virginia Tech, Blacksburg, Virginia 24061, USA}
\author{Vito W. Scarola}
\affiliation{Department of Physics, Virginia Tech, Blacksburg, Virginia 24061, USA}
\author{Brian DeMarco}
\email{bdemarco@illinois.edu}
\affiliation{Department of Physics,  University of Illinois at Urbana-Champaign, Urbana, Illinois 61801, USA}
\date{\today}

\begin{abstract}
We report the realization of correlated, density-dependent tunneling for fermionic $^{40}$K atoms trapped in an optical lattice.   By appropriately tuning the frequency difference between a pair of Raman beams applied to a spin-polarized gas, simultaneous spin transitions and tunneling events are induced that depend on the relative occupations of neighboring lattice sites.  Correlated spin-flip tunneling is spectroscopically resolved using gases prepared in opposite spin states, and the inferred Hubbard interaction energy is compared with a tight-binding prediction.  We show that the laser-induced correlated tunneling process generates doublons via loss induced by light-assisted collisions.  Furthermore, by controllably introducing vacancies to a spin-polarized gas, we demonstrate that correlated tunneling is suppressed when neighboring lattice sites are unoccupied.
\end{abstract}
\maketitle

Measurements on ultracold atoms trapped in optical lattices have emerged as a powerful approach to studying quantum phase transitions and dynamics in strongly correlated systems.  Periodic driving forces and light-induced tunneling combined with optical lattices have enabled experiments to achieve physics beyond the minimal Hubbard model (see Ref. \citenum{RevModPhys.89.011004} for a recent review).  For example, magnetic phase transitions have been probed \cite{parker2013direct,struck2011quantum}, synthetic gauge fields realized \cite{PhysRevLett.102.130401,PhysRevLett.111.185301,PhysRevLett.111.185302,PhysRevLett.108.225304}, and non-trivial band structures \cite{Jotzu2014} have been created using periodic driving and external fields in lattices.

In this work, we use applied laser fields to demonstrate correlated tunneling that depends on density and spin for fermionic atoms.  Correlated tunneling, known in solids as a bond-charge interaction, has been proposed to play a role in high-temperature superconductivity \cite{hirsch1989bond} and lattice stiffening in polyacetylene \cite{PhysRevLett.42.1698,PhysRevB.42.475}.  The influence of correlated tunneling on transport properties has also been intensively investigated in quantum dots, where it can be manipulated by gate voltages and applied electromagnetic fields \cite{van2002electron}.  Beyond mimicking these effects in optical lattices, correlated tunneling for ultracold atoms has attracted theoretical interest for inducing occupation-dependent gauge fields \cite{greschner2014density}, obtaining novel phases such as holon and doublon superfluids \cite{rapp2012ultracold}, and realizing anyonic Hubbard models \cite{greschner2015anyon}.  Thus far, density dependent tunneling has been observed for bosonic atoms trapped in optical lattices via lattice modulation \cite{ma2011photon, meinert2016floquet}.

Inspired by the theoretical proposals in Ref.~\citenum{bermudez2015interaction}, we implement a new experimental approach to generate spin and density-dependent tunneling for fermionic atoms.   We apply a pair of Raman beams to a spin-polarized gas, for which conventional tunneling is forbidden by the Pauli exclusion principle, to flip the atomic spins and induce density-dependent tunneling. We spectroscopically resolve  correlated spin-flip tunneling (CSFT) events, and the corresponding increase in doubly occupied sites is measured using loss from light-assisted collisions.  Moreover, by varying the filling fraction in the lattice, we directly verify the density-dependence of spin transitions.

These measurements are performed using a degenerate Fermi gas composed of $^{40}$K atoms trapped in a cubic optical lattice in a regime described by a single-band Fermi-Hubbard model with tunneling energy $t$ and interaction energy $U$.  Overall confinement is provided by a 1064~nm optical dipole trap.  After cooling the gas in the dipole trap and before slowly superimposing the lattice, the gas is spin polarized in either the $\ket{F=9/2,m_F=9/2}$ or $\ket{F=9/2,m_F=7/2}$ state, which we label $\ket{\uparrow}$ and $\ket{\downarrow}$.  The atom number and confinement are tuned so that the central density is approximately one atom per site, with the Fermi energy $E_F\approx 7 t$.  A pair of Raman beams with wavevectors $\vec{k}_1$ and $\vec{k}_2$ intersecting at approximately $30^\circ$ are focused onto the gas and pulsed to drive spin transitions (Fig. 1).  The frequency difference $\Delta\omega=\omega_1-\omega_2$ between the Raman beams is tuned near to the $\ket{\uparrow}\rightarrow\ket{\downarrow}$ resonance.  The number of atoms in each spin state after a Raman pulse is measured using a magnetic field gradient applied during time-of-flight imaging.

\begin{figure}[!htp]
    \includegraphics[width=\linewidth]{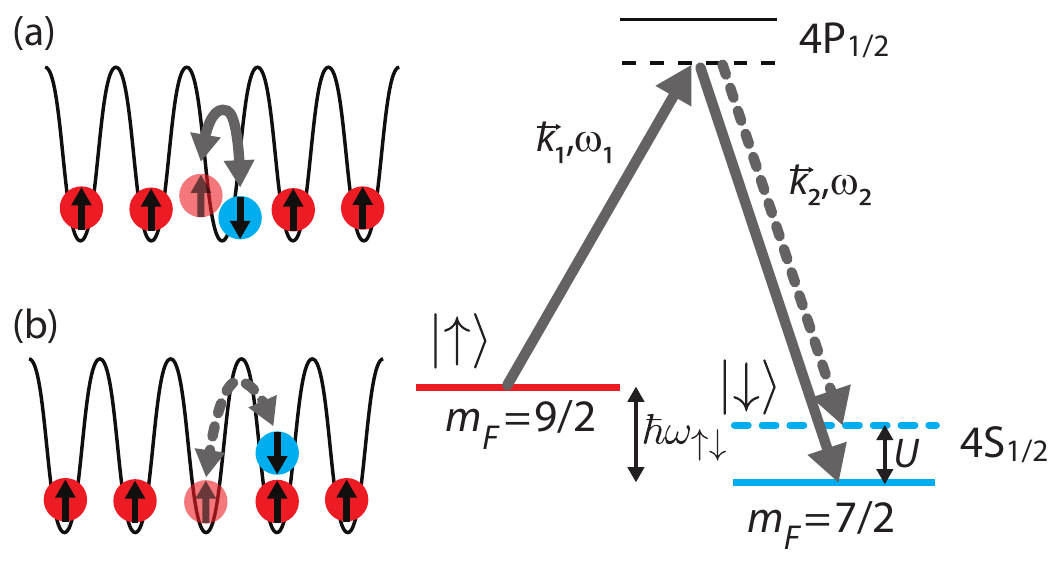}
      \caption{Schematic diagram of Raman transitions.  A pair of Raman beams with frequencies $\omega_{1,2}$ and wavevectors $\vec{k}_{1,2}$ is applied to drive transitions between the $\ket{\uparrow}$ (red) and $\ket{\downarrow}$ (blue) states.  The Raman wavevector difference $\vec{\delta k}=\vec{k}_1-\vec{k}_2$ lies along the $\left(-1,-1,1\right)$ direction of the lattice.  Selecting between two distinct processes is achieved by fixing the laser beam frequency $\omega_1$ and tuning $\omega_2$.  (a) If the frequency difference matches the Zeeman energy (i.e., $\Delta\omega=\omega_1-\omega_2=\omega_{\uparrow\downarrow}$), then atoms flip their spin and remain on the same site.  (b)  When the laser frequency difference accommodates the Hubbard interaction energy $U$ ($\Delta\omega=\omega_{\uparrow\downarrow}-U/\hbar$), then CSFT occurs and atoms tunnel to neighboring occupied sites and flip their spin.  For $\ket{\downarrow}$ as an initial state, the condition for resonant CFST changes to $\Delta\omega=\omega_{\uparrow\downarrow}+U/\hbar$.}
      \label{fig:cartoon}
\end{figure}

The Raman beams can drive two resonant processes depending on $\Delta\omega$.  If $\Delta\omega$ is tuned to the energy difference between spin states ($\Delta\omega=\omega_{\uparrow\downarrow}$), then on-site spin rotations occur without induced tunneling and changes in site occupancies (Fig. 1a). We define this process as the carrier transition.  By tuning the frequency difference between the beams to include $U$ ($\Delta\omega-\omega_{\uparrow\downarrow}=\pm U/\hbar$), density-dependent tunneling is driven as a sideband to the carrier (Fig. 1b). Working in the large $U$ limit, we use perturbation theory to find the effective CSFT Hamiltonian for this process:
\begin{multline}
H_{\rm CSFT}= \sum_{ i,j \in\text{n.n.}}\left[
K_{ij}n_{i\uparrow}(1-n_{j\downarrow})\hat{c}_{j\uparrow}^{\dagger}\hat{c}^{\vphantom{\dagger}}_{i\downarrow}\right.+\\
\left.K^{*}_{ij}n_{i\uparrow}(1-n_{j\downarrow})\hat{c}_{i\downarrow}^{\dagger}\hat{c}^{\vphantom{\dagger}}_{j\uparrow}\right]
\end{multline}
(see Supplemental Material \cite{SM} for theoretical details).  Here, $i,j \in\text{n.n.}$ denotes a sum over nearest neighboring sites and permutations, $\hat{c}^\dagger_{i\sigma}$ ($\hat{c}_{i\sigma}$) creates (removes) a particle on site $i$ in spin state $\sigma$, and $n_{i\sigma}$ is the number of particles on site $i$ in state $\sigma$. Eq. 1 has been projected into the subspace connected to the initial spin-polarized $\ket{\uparrow}$ state by resonant CSFT.  We have verified that the dynamics of the full tight-binding CSFT model are reproduced by Eq. 1 by applying the time-evolving block decimation algorithm to 1D chains \cite{vidal2007classical,SM}.

CSFT arises as a spin-flip transition to a virtual state offset by $U$ followed by a tunneling event.  In contrast to the conventional tunneling term $-t \hat{c}^\dagger_i \hat{c}_j$ in the Fermi-Hubbard model, this laser-induced correlated spin-flip tunneling is density dependent and accompanied by a spin rotation.  CSFT occurs only when neighboring sites are occupied by atoms in the same spin state or when a doublon (i.e., a $\ket{\uparrow}$--$\ket{\downarrow}$ pair) is next to an empty site.  The magnitude of the CSFT matrix element $\left|K_{ij}\right|\approx \left|\Omega\left( 1-e^{i \vec{\delta k}\cdot \vec{d}}\right)\right| t/2 U$, where the ratio $t/U\approx 0.04$--0.08 is controlled by the lattice potential depth $s$, $\vec{\delta k}$ is the Raman wavevector difference, $\vec{d}$ is a lattice vector, and $\vec{\delta k}\cdot \vec{d}=\pi/2\sqrt{3}$ for all lattice directions in our experiment.  The Rabi rate for the carrier transition $\left|\Omega\right|\approx 2\pi\times650~$Hz is controlled by the Raman laser intensity and measured via Rabi oscillations \cite{SM}.

The site-dependent Raman phase $e^{i \vec{\delta k}\cdot\vec{d}}$ that arises from the Raman wavevector difference is critical to allow tunneling to occur.  When spin rotations are driven by long-wavelength radiation or co-propagating Raman beams, this phase factor is absent, and tunneling is prevented.  One reason for the absence of tunneling in this scenario can be understood as destructive interference between multiple tunneling pathways caused by the anti-symmetrization of the fermionic wavefunctions \cite{SM}.  This effect is related to the behavior of clock shifts for fermionic atoms \cite{gupta2003radio,PhysRevLett.103.260402,PhysRevLett.110.160801}. In our case, the Raman phase factor suppresses the destructive tunneling interference.

We spectroscopically resolve CSFT and distinguish it from on-site spin rotations by measuring the change in spin fraction after a Raman pulse.  Sample data are shown in Fig. 2a for the fraction $f_{\downarrow,\uparrow}$ of atoms transferred between spin states at varied $\Delta\omega$ for $s=10$~$E_R$, where $E_R=h/2md^2$ is the recoil energy, $m$ is the atomic mass, and $d\approx390$~nm is the lattice spacing.  To observe the relatively slow CSFT process \cite{SM}, the Raman pulse is $50$~ms long.  The carrier transition is therefore over-driven, which results in a broad feature that obscures CSFT---when $\Delta\omega-\omega_{\uparrow\downarrow}\approx\pm U/\hbar$, atoms undergo both detuned carrier spin-flip transitions and resonant CSFT.  To overcome this complication and isolate CSFT, we subtract the data taken at identical $\Delta\omega$ with opposite initial spin configurations.  Since the carrier frequency does not depend on the initial state, the contribution from the broad carrier feature is cancelled out by this procedure.  In contrast, the frequency offset of the CFST sideband changes sign with the initial spin configuration and is not removed by the subtraction.  The resulting lineshape for $f_\uparrow-f_\downarrow$ shown in Fig. 2b therefore reveals the CFST sidebands as peaks offset at approximately $\pm U\approx\pm3.5$~kHz from the carrier transition.

To compare with the predicted sideband frequency, we fit the $f_\uparrow-f_\downarrow$ lineshape to a sum of two gaussian functions with independent central frequencies and standard deviations as free parameters.   The interaction energy $U$ determined from this fit as half of the frequency separation of the peaks is shown in Fig. 2c for data taken at different lattice potential depths.  The inferred $U$ increases less rapidly with $s$ than the tight-binding prediction, which is shown as a dashed line.  Renormalization of $U$ by the Raman process \cite{SM} and by contributions from higher bands \cite{lacki2013} may contribute to this discrepancy.

\begin{figure}[!htp]

    \includegraphics[width=\linewidth]{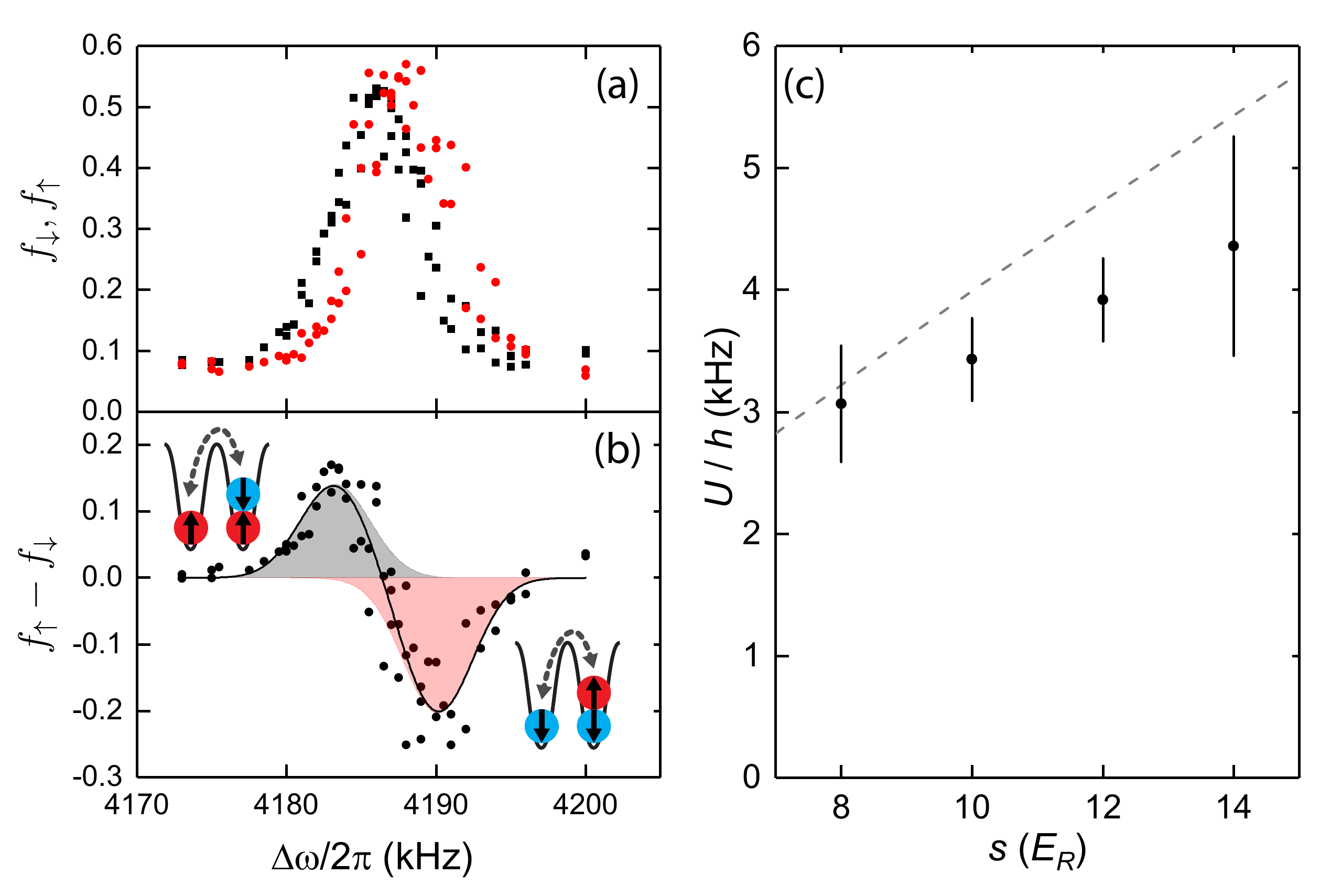}
      \caption{Spectroscopy of CSFT.  (a) The fraction of atoms transferred between spin states by a 50~ms Raman pulse is shown for an initially $\ket{\uparrow}$ (black squares, $f_\uparrow$) and $\ket{\downarrow}$ (red circles, $f_\downarrow$) spin polarized state at $s=10$~$E_R$ for varied $\Delta\omega$.  For these measurements, $N=25400 \pm 3900$ atoms were cooled to $T/T_F=0.24 \pm 0.08$ before turning on the lattice.  (b) The difference $f_\uparrow-f_\downarrow$ for pairs of points in (a) reveals the CSFT sidebands at approximately $\pm U$.  The black line shows a fit to a sum of two gaussian functions; the individual gaussians are displayed as shaded regions.  The peak at lower (higher) frequency corresponds to CSFT for an initially $\ket{\uparrow}$ ($\ket{\downarrow}$) spin-polarized state. (c) The interaction energy $U$ inferred from fits to data such as those shown in (b) for varied $s$.  The error bars are derived from the fit uncertainty.  The dashed line is the value of $U$ from a standard tight-binding calculation.}
      \label{fig:dif}
\end{figure}

Each CSFT event for a spin-polarized gas is accompanied by the creation of a doublon--hole pair.  We probe doublon generation using loss induced by light-assisted collisions (LAC) \cite{depue1999unity}.   The carrier frequency $\omega_{\uparrow\downarrow}$ is estimated using a gaussian fit to Raman spectroscopy taken using a 0.7~ms pulse, which is too short to drive CSFT.  The inset to Fig. 3 shows sample LAC data taken after a 50~ms Raman pulse with $\left(\Delta\omega-\omega_{\uparrow\downarrow}\right)/2\pi=3.5$~kHz, which corresponds to the $+U$ CSFT sideband. Immediately following the Raman pulse, the lattice potential depth is rapidly increased to 29~$E_R$ to arrest further dynamics.   We measure the number of atoms remaining after a laser pulse 50~MHz detuned from the $4S, F=9/2\rightarrow 5P_{3/2}, F=11/2$ transition is applied to the gas in the 29~$E_R$ lattice.  Two loss processes are evident as the duration $\tau$ of the resonant laser pulse is changed.  The loss on a fast timescale $\tau_D$ corresponds to LAC removing atoms from doubly occupied sites, while the decay over a slower timescale $\tau_S$ results from single atoms ejected from the dipole trap via spontaneous scattering.  These data are fit to a double exponential decay function $N\left(\tau\right)=N_D e^{-\tau/\tau_D}+\left(N-N_D\right) e^{-\tau/\tau_S}$ with $N$, $N_D$, $\tau_D$ and $\tau_S$ as free parameters to determine the fraction of doubly occupied sites $D=N_D/N$.

Measurements of $D$ as $\Delta\omega$ is changed show that a resonance for doublon creation is centered near the CSFT spectroscopy sideband peak at $\left(\Delta\omega-\omega_{\uparrow\downarrow}\right)\approx U/\hbar$ (Fig. 3). The data are compared with the fit from Fig. 2b to $\left|f_\uparrow-f_\downarrow\right|$, which can be interpreted as the fraction of atoms that flip their spin during CFST events.  The close agreement between the fit and $D$ imply that each spin-flip is associated with the creation of a doubly occupied site.

\begin{figure}[!htp]
    \includegraphics[width=\linewidth]{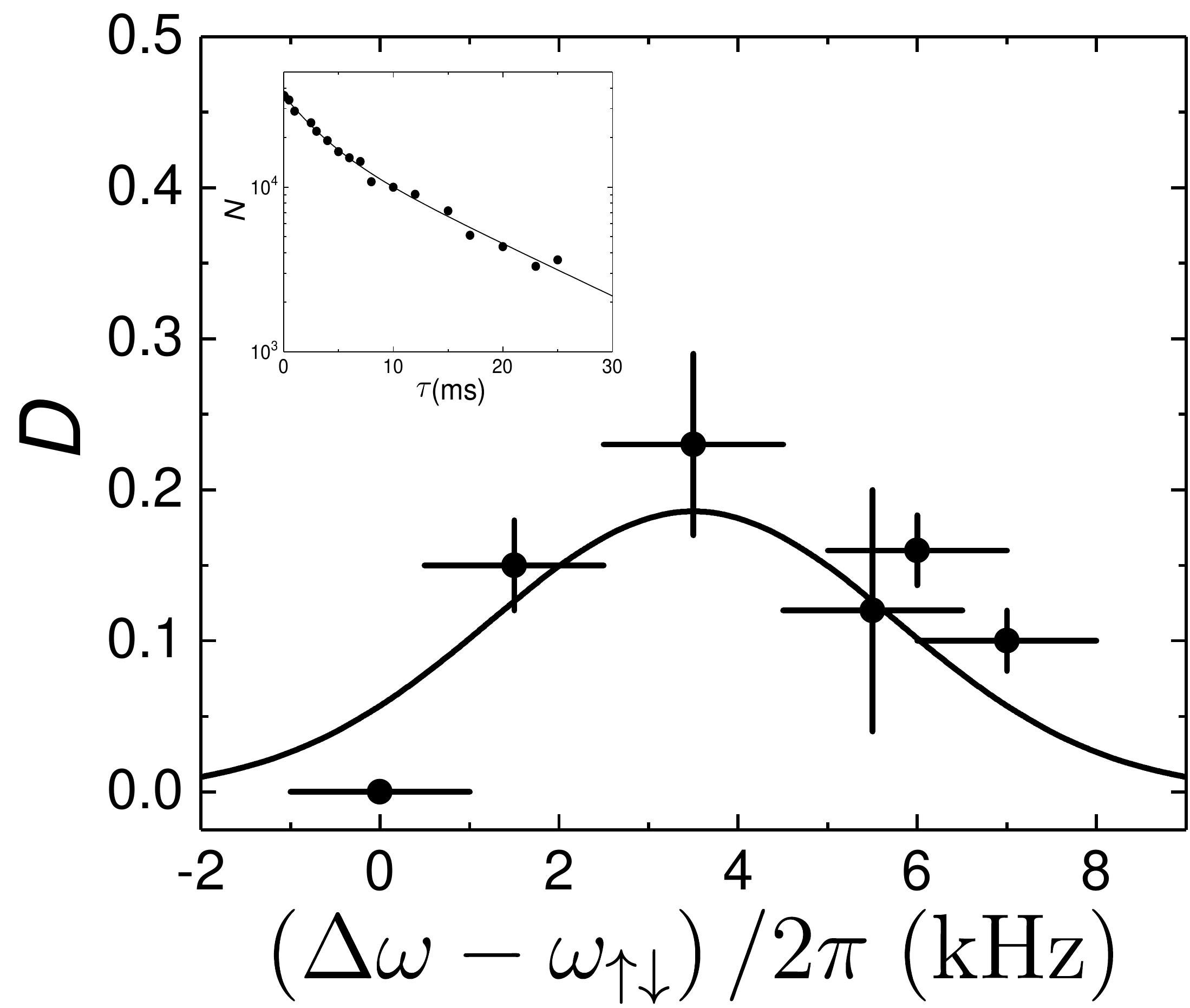}
      \caption{Fraction of doubly occupied sites $D$ measured after a 50~ms Raman pulse at various detunings for $s=10$~$E_R$.  The inset shows sample LAC for $\left(\Delta\omega-\omega_{\uparrow\downarrow}\right)/2\pi=3.5$~kHz fit to a double-exponential decay with $\tau_D=3.1\pm 0.77$~ms and $\tau_S= 13.7\pm 2.5$~ms.  The vertical error bars in $D$ are derived from fits to similar data acquired at different $\Delta\omega$, while the horizontal error bars show the estimated 0.5~kHz uncertainty in the carrier transition from magnetic field drift.  The solid line in the main panel is the fit from Fig. 2a for $\left|f_\uparrow-f_\downarrow\right|$ plotted on the same scale as $D$.}
      \label{fig:FracLAC}
\end{figure}

Finally, we demonstrate the sensitivity of CSFT to site occupancy by reducing the atom number and controllably introducing vacancies before a Raman pulse on the CSFT sideband (see Fig. 4 inset).  Our technique involves three steps.  After turning on the lattice to $s=8$~$E_R$, atoms are transferred from the $\ket{\uparrow}$ to the $F=7/2, m_F=7/2$ state via adiabatic rapid passage (ARP) driven by a microwave-frequency magnetic field.  The power of the microwave field (swept across 0.4~MHz in 0.5~ms) is varied to control the probability of a transition between hyperfine states.  The fraction $\delta N$ of atoms remaining in the $\ket{\uparrow}$ state are removed from the lattice with a 0.5~ms pulse of light resonant with the $4S,F=9/2\rightarrow 5P_{3/2}, F=11/2$ transition.  A second ARP sweep (across 0.8~MHz in 1~ms) returns all of the atoms shelved in the $F=7/2$ manifold to the $\ket{\uparrow}$ state.

After this procedure, unoccupied sites are randomly distributed through the spin-polarized atomic density distribution.  The presence of holes suppresses CSFT, which can only occur when adjacent sites are occupied.  We probe this effect by measuring changes in $\left|f_\uparrow-f_\downarrow\right|$ for a 40~ms Raman pulse with $\Delta\omega$ fixed on the $\pm U$ peaks of the CSFT sideband (Fig. 4).  As the number of atoms is reduced and the hole density increases, $\left|f_\uparrow-f_\downarrow\right|$ decreases, indicating that fewer atoms can participate in CSFT.  The data shown in Fig. 4 show good agreement with a prediction of the probability for adjacent sites to be occupied \cite{SM}.  For this calculation, the density profile is computed using entropy matching based on $s$, the overall confinement, and the measured $N$ and $T/T_F$.  The probability of adjacent sites being occupied is determined by averaging over configurations that involve randomly removing a fraction $\delta N$ of atoms from the simulated density profile.

\begin{figure}[!htp]
    \includegraphics[width=\linewidth]{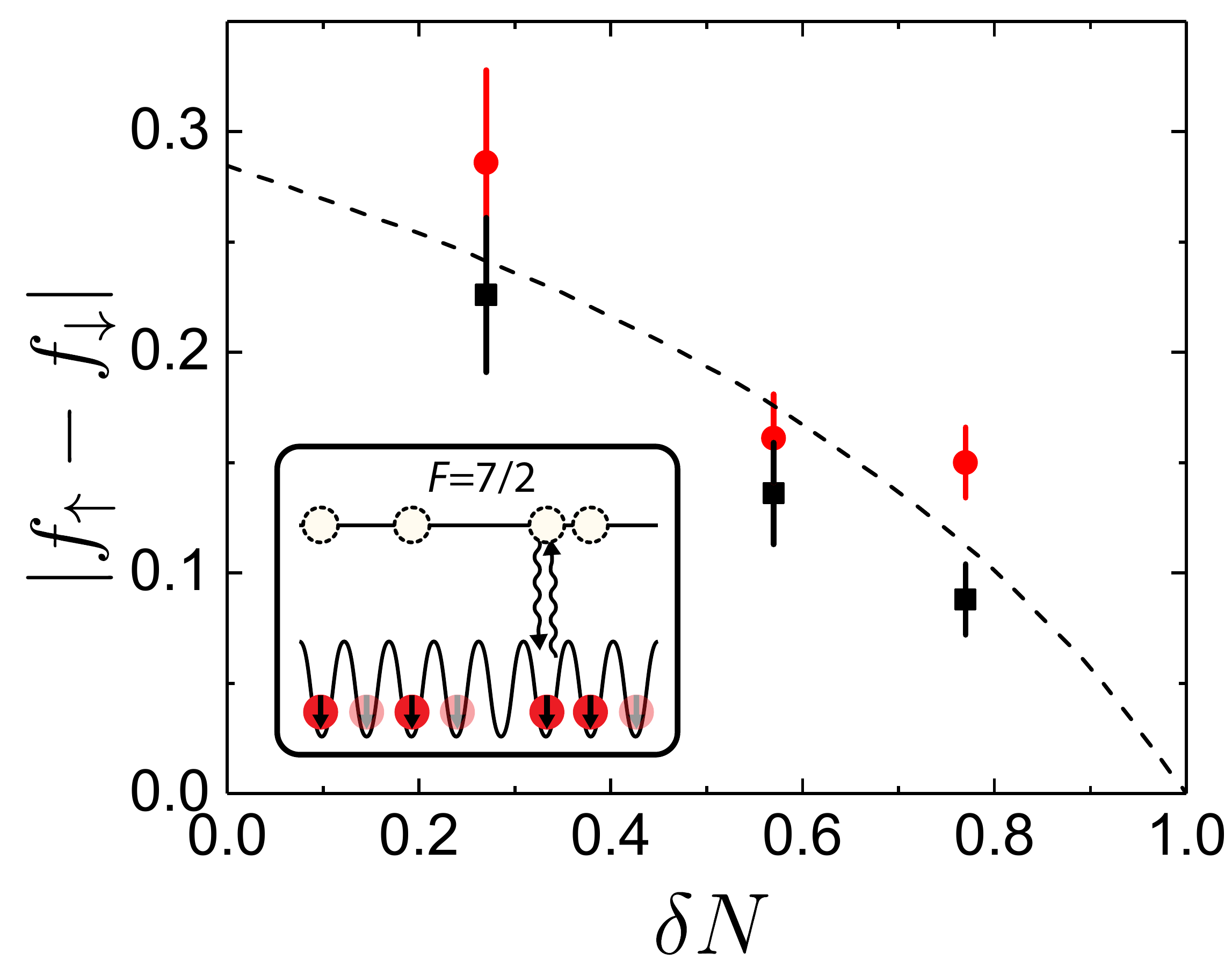}
      \caption{Density dependence of CSFT.  The CSFT spectroscopy signal taken with fixed $\left(\Delta\omega-\Delta\omega_{\uparrow\downarrow}\right)\approx \pm U/\hbar$ is shown for varied fraction $\delta N$ of atoms randomly removed from an $s=8$~$E_R$ lattice gas.  For these data, $N=47000$--81000, and the gas was cooled to $T/T_F\approx 0.35$ before turning on the lattice.  Data obtained with the $+U$ sideband are shown as red circles and those for $-U$ as black squares.  The sideband frequencies were determined using a double-gaussian fit to CSFT spectroscopy data, as in Fig. 2b.  The dashed line is a prediction for the probability to find adjacent sites occupied based on a calculation of the density profile after the removal procedure.  The inset shows the procedure for controllably introducing vacancies. Atoms (shown as transparent) that are not shelved in the $F=7/2$ state via microwave transitions are removed using resonant light.}
      \label{fig:VarDensity}
\end{figure}

In conclusion, we have reported the first observation of density-dependent tunneling in an optical lattice Fermi-Hubbard model.  In the future, the technique we have developed may be used to observe exotic states such as bond-ordered waves, triplet pairing, and hole superconductivity \cite{bermudez2015interaction}.  The site-dependent phase of the bond-charge interaction the Raman lasers introduce also leads to an synthetic gauge field that was not explored in this work.  The unique properties of the occupation-dependent gauge field created via this method can be used to simulate interacting relativistic quantum field theories and correlated topological insulators \cite{bermudez2015interaction}.

\acknowledgments{We acknowledge support from the Army Research Office (W911NF-17-1-0171) and National Science Foundation (PHY 15-05468).}

\bibliography{RefDoublon}
\end{document}